\documentclass[aps,prl,showpacs,twocolumn,superscriptaddress]{revtex4}

\usepackage{epsfig}
\usepackage{graphicx}
\usepackage{amsmath}
\usepackage{amssymb}
\usepackage{amsfonts}
\usepackage{dcolumn}
\usepackage{color}

\begin{document}
\title{High efficiency preparation of single trapped atoms using blue detuned light assisted collisions
}

 \author{A. V. Carpentier}
 \affiliation{Jack Dodd Centre for Quantum Technology, Department of Physics, \\ University of Otago, New Zealand.}
\author{Y. H. Fung}
 \affiliation{Jack Dodd Centre for Quantum Technology, Department of Physics, \\ University of Otago, New Zealand.}
\author{P. Sompet}
 \affiliation{Jack Dodd Centre for Quantum Technology, Department of Physics, \\ University of Otago, New Zealand.}
 \author{A. J. Hilliard}
 \affiliation{Jack Dodd Centre for Quantum Technology, Department of Physics, \\ University of Otago, New Zealand.}
 \author{T. G. Walker}
 \affiliation{Department of Physics, University of Wisconsin-Madison, Madison, Wisconsin 53706.}
 \author{M. F. Andersen}
 \affiliation{Jack Dodd Centre for Quantum Technology, Department of Physics, \\ University of Otago, New Zealand.}

\begin{abstract}
We report on a procedure by which we obtain a  $91\%$ loading efficiency of single $^{85}$Rb atoms in an optical microtrap.
This can be achieved within a total preparation time of  $542$ ms.
We employ blue detuned light assisted collisions to realize a process in which only one of the collision partners is lost.
We explain the mechanism for efficiently loading a single atom and discuss the factors that limit the final efficiency.
\end{abstract}

\maketitle
The ability to control microscopic systems drives the development of a wide range of modern technologies. Recent years have witnessed tremendous
progress towards complete control of individual atoms \cite{Blatt,Wineland,Bloch,arno}.
This is of strong interest due to their potential use as qubits in a quantum logic device \cite{Blatt,Wineland,Feynman,Divicenzo,Jessen}.
Neutral atoms in far off-resonance optical microtraps have several favorable properties for quantum information processing: a qubit can
be encoded into the internal states of each atom, the atoms are well shielded from the environment, and long range entangling
interactions can be switched on and off \cite{Pzoller,Saffman,Wilk}.

Several schemes for quantum gate operations using cold neutral atoms have been investigated experimentally \cite{Anderlini,Saffman,Wilk}. Two qubit entangling operations have been demonstrated using Rydberg mediated dipolar interactions \cite{Saffman,Wilk}. Operations involving more than two qubits require an effective way to deterministically prepare single
trapped atoms that can be individually addressed. To this end, multiple schemes have been investigated. The use of degenerate
gases for creating Mott insulators \cite{Greiner,Bloch} or the isolation of fermions in single trap geometries \cite{serwane}
lead to complicated experiments with long loading times.
The Rydberg blockade mechanism has been proposed as a fast deterministic loading scheme, but this approach has yet to be demonstrated experimentally \cite{Beterov}.
Finally, a range of experiments has used light assisted collisions.
In experiments that employed red detuned light to induce the collisions, the loading efficiency has been limited to $50\%$ \cite{DePueMT,Schlosser}.
Our previous work showed that a scheme using blue detuned light could increase the loading efficiency to $83\%$ \cite{andersen}.
However, further improvement was still required to load quantum registers of $30$ qubits  without the need for consecutive atom sorting \cite{arno}. Thirty qubits is considered a lower bound for quantum computing to exceed the limitations of classical computing  \cite{quantumRegister}.

In this Letter we report a loading efficiency of single $^{85}$Rb atoms of $91\%$. The loading procedure can be completed within $542$ ms.
We achieve this by employing blue detuned light assisted collisions to realize a process in which only one of the collision partners is lost.
The process relies on a combination of laser cooling to remove excess energy, and controlled transfer of energy
through the light assisted collisions. We explain the mechanism and identify the factors that still limit the loading efficiency.

%
%%%%%%%%%%%%% figure 1 %%%%%%%%%%%%%%
\begin{figure}
 \epsfig{figure=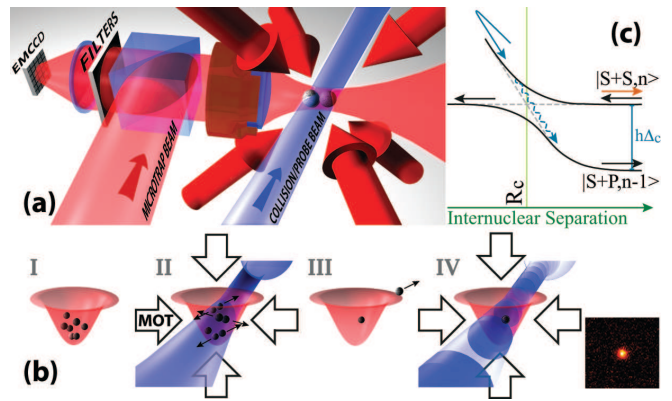, width=1\columnwidth}
\caption{(Color online)  (a) Sketch of the experimental setup.
(b) Experimental process: (I) $^{85}$Rb atoms are loaded into the optical microtrap;
(II) the blue detuned collision beam induces inelastic light assisted collisions while the MOT cooling beams cool the atoms;
(III) when only one atom remains the expelling process ends; (IV) a blue detuned standing wave together with the MOT cooling beams
are used during the imaging stage.
(c) One-dimensional dressed state picture of light-assisted collisions. The different arrows represent:
black (up, right): two atoms in ground state approaching; black (left): LZ transition; black (low): inelastic collision; orange (up, right): elastic
collision; continuous blue (up, left): adiabatic transition; dashed blue: LZ transition.
}
\label{fig1}
\end{figure}
%%%%%%%%%%%%%%%%%%%%%%%%%%%%%%%
%
The advantage of using blue detuned light for inducing collisions is that it allows us to control the energy
released in each inelastic collision. This can be understood through a simplified two-level model.
We consider it in the dressed state picture illustrated in Fig. \ref{fig1}(c) \cite{wiener},
where the $|S+S\textgreater$ asymptote represents the two atoms in their ground states, offset by one photon energy, and $|S+P\textgreater$
represents one atom in the ground state and one in the excited state.
The energy of the $|S+P\textgreater$ state is below that of $|S+S\textgreater$ when the light is blue detuned and above when it is red detuned.
For the purpose of this experiment, we approximate the $|S+P\textgreater$ curve with its  $C_3/R^3$ asymptote (with $R$ the interatomic distance and $C_3$ a constant), while the $|S+S\textgreater$ curve is considered independent of $R$.
As the resonance condition depends on the internuclear separation, the dressed states will cross each other at the Condon point $R_c$
in the absence of coupling. This becomes an avoided crossing in the presence of light.
When two atoms pass the Condon point
they can undergo two different processes: a Landau-Zener (LZ) transition to the other dressed state, or they can
move adiabatically through the coupling region while remaining in the same dressed state \cite{wiener}.
In a collision process the atoms cross $R_c$ twice: when the atoms approach, and again
when they move away from each other.
For the atoms to undergo an inelastic light assisted collision, they must undergo an adiabatic passage followed by a LZ transition,
or vice versa, ending, in either case on the $|S+P\textgreater$ asymptote.
By this process, the atom pair gains an energy $h\Delta_c$ equal to the difference in dressed state energies at large interatomic distance. 
The probability for this to occur when $R$ reach $R_c$ is $P_I=2P_{LZ}\left(1-P_{LZ} \right)$,
where $P_{LZ}=\exp(\frac{-2\pi\hbar\Omega^2R_c^4}{3C_3v})$ is the probability for a LZ transition, with $\Omega$ the on-resonance Rabi frequency and $v$ the relative radial speed of the atom pair at $R_c$. As will be explained below, balancing the collisional energy transfer with laser cooling allows preferential ejection of single atoms from the trap.  Thus, the atoms are lost one by one until only a single atom remains and the collisions cease.

Figure \ref{fig1}(a) shows a sketch of the experimental setup.
A high numerical aperture aspheric lens mounted inside the vacuum chamber focuses
a far off-resonance ($\lambda=828$ nm) laser beam with power $30$ mW to a waist of $1.8$ $\mu$m.
The resulting $U_0=h\times85$ MHz deep optical microtrap is loaded with $^{85}$Rb atoms using a
magneto-optical trap (MOT) followed by a compressed MOT and  an optical molasses stage.
The sequence loads on average $10$ to $80$ atoms depending on the MOT stage duration.

Figure \ref{fig1}(b) illustrates how we isolate and detect a single atom from the initial ensemble.
With the MOT beams reconfigured for cooling the atoms in the microtrap, we add a blue-detuned ``collision beam'' of controllable pulse duration. 
The collision beam, of power $11$ $\mu$W, has a radius of $\sim150$ $\mu$m at the position of the microtrap
and is blue detuned by $\Delta_c$ from the $D1$ $F=2$ to $F'=3$ transition at the center of the trap.
After the collision pulse, the final step is to take an image of the atomic sample to measure the atom number.
The method for imaging and counting the atoms is described in \cite{andersen2}. We have a single atom detection efficiency of $99.5\%$.

%
%%%%%%%%%%%%% figure 2 %%%%%%%%%%%%%%
\begin{figure}
 \epsfig{figure=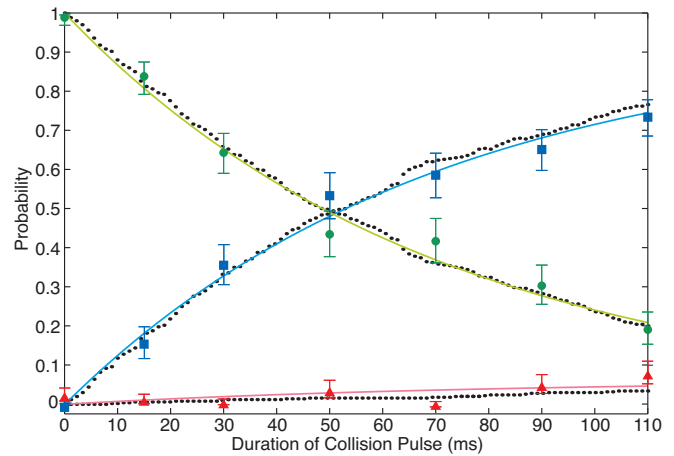, width=1\columnwidth}
\caption{(Color online)  Atom pair's evolution as a function of the collision light pulse duration.
The green circles, blue squares and red triangles
show the probabilities of two, one, or zero atoms respectively after the collision pulse. The
solid lines are a fit of the experimental data. The pair decay time is $\sim70$ ms. The dotted lines
show the result of a simulation of the experiment. 
Error bars represent throughout the paper a $68.3\%$ confidence interval based on binomial statistics. 
}
\label{fig2}
\end{figure}
%%%%%%%%%%%%%%%%%%%%%%%%%%%%%%%
%
%

The collision stage (frame $II$ and $III$ in Fig. \ref{fig1}(b)) determines the efficiency with which a single atom is prepared.
Two mechanisms can reduce this to a value below the ideal $100\%$. First: as a single atom can be prepared before the end of the
collision stage, any process leading to its subsequent loss will reduce the probability of retaining the atom until imaging.
Second: collisional processes by which both partners are lost reduce the efficiency as the final pair can be lost together,
again leading to no trapped atoms at the time of detection.

The mechanism for inelastic collisions where only one of the collision partners is lost, relies on the release of sufficient energy such that, after
a collision, the pair has enough energy for one atom to be lost but not enough for both to leave the trap. At the same time, the pair should share the
total energy unevenly, such that one of the atoms has a sufficient fraction of the pair's energy ($E_p$) to escape the trap.
The pair should therefore have significant center of mass motion before the collision. If $E_p$ before a collision is larger than zero but less than the trap depth, and the inelastic collision releases an energy equal
to the trap depth, then there
is a finite probability for one atom to be lost, but the probability that both are lost is zero. In our
experiments, we aim to favor inelastic collisions in this regime. 

For the method to be efficient we need a mechanism to suppress the
following undesirable scenario:
as the center of mass motion of the pair is not controlled in detail,
collisions can occur such that $U_0\textless E_p\textless 2U_0$,
but with each atom having $E\textless U_0$ so that neither is lost. A second inelastic collision will now leave the pair
with enough energy for both atoms to escape. It is therefore important to remove some of the pair's energy between
inelastic collisions. This is the primary role of the spectrally reconfigured MOT cooling beams during the collisions in stage $II$ of Fig. \ref{fig1}(b). 
In this stage, each of the MOT beams has a peak intensity of $I_M=35~\mathrm{W/m^2}$ and are 
$4$ MHz red detuned from the $D2$ $F=3$ to $F'=3$ transition for atoms at the center of the trap (averaged across the magnetic sublevels).
By preparing single atoms, heating them, exposing them to a collision/cooling light pulse of variable duration,
and subsequently measuring their temperature using the release recapture method \cite{ReleaseRecapture},
we have measured the temperature evolution of single atoms exposed to collision and cooling light. The data is well fitted by an exponential decay with time constant of $5$ ms to an equilibrium temperature of $\sim250$ $\mu$K.
In addition to cooling, the MOT beams provide efficient optical pumping of atoms into the desired $F=2$ ground state in the center of the trap where collisions are likely to happen.

The inelastic collision rate drops with increasing $E_p$ such that the cooling provided by the MOT beams typically can lower the energy of pairs with $E_p\textgreater U_0$ before another inelastic collision occurs. This happens because the atomic density drops with increasing $E_p$, and because we operate our collision beam in a parameter regime where $P_I$ drops with $v$.

To investigate the single atom ejection process we have loaded pairs of atoms into the microtrap and
measured the number of atoms left after a collision/cooling pulse. Figure \ref{fig2} shows the probability of measuring
zero, one, and two atoms as a function of the collision pulse duration using $11$ $\mu$W of
collision beam power with a detuning of $\Delta_c=85$ MHz. The experimental data is corrected for offsets due to loss during detection.
The dotted lines in Fig. \ref{fig2} show a simulation of our experiment based on the above description of the process.
In the simulation, the motion of two atoms in the Gaussian potential is treated classically. A Doppler cooling model \cite{dopplerCool}, with parameters
adjusted to reproduce the measured temperature evolution of single atoms, damps the motion of energetic trapped atoms. When the distance between the atoms reaches the Condon radius they may undergo an inelastic collision with probability $P_I$. The collision
conserves center of mass momentum while releasing an energy of $h \Delta_c$. When an atom reaches an energy above the trap depth it is
considered lost. The probabilities are generated from 500 pairs and show good agreement with the experiment. In order to compensate for the use of the simple two-level model in predicting $P_I$, we adjusted the Rabi-frequency in $P_{LZ}$ such that the decay time of pairs matches the experimental value. Using a collision beam power of $7$ $\mu$W and $\Delta_c=185$ MHz we repeated the measurement and simulation of Fig. \ref{fig2}. Similar agreement was found for these parameters, and the adjustment factor for the Rabi-frequency matched that from Fig. \ref{fig2} within 7.5 $\%$.

%
%%%%%%%%%%%%% figure 3 %%%%%%%%%%%%%%
\begin{figure}
 \epsfig{figure=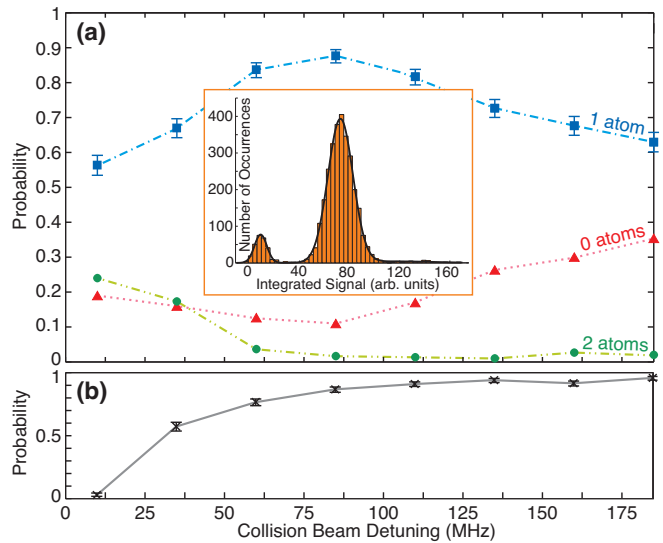, width=1\columnwidth}
\caption{(Color online)  Inset: Histogram of the integrated fluorescence for $3200$ realizations  of the experiment.
The large peak represents the single atom loading efficiency and comprises the $91\%$ of the realizations.
(a): Probabilities versus detuning of the collision beam ($\Delta_c)$. Symbols are as in Fig. \ref{fig2}.
(b): Single atom survival probability after a collision pulse of $3.5$ s, which is a measure of the single atom lifetime.
 }
\label{fig3}
\end{figure}
%%%%%%%%%%%%%%%%%%%%%%%%%%%%%%%
%

The inset of Fig. \ref{fig3} shows  the histogram of the integrated fluorescence for $3200$
realizations of the experiment using the parameters of Fig. \ref{fig2}. The single atom loading efficiency of $91\%$ is represented by the large peak.
The total time required to do this is $542$ ms ($50$ ms MOT, $102$ ms CMOT, $5$ ms molasses to load on average
$19$ atoms into the microtrap, and a $385$ ms collision pulse). Reducing the MOT loading time to $25$ ms lead to events of no atoms loaded into the microtrap before the collision pulse.
To illustrate the potential of the method, we consider the time it would take to load an array of 30 microtraps in parallel. The probability for all 30 sites to be occupied with a single atom is $\sim 0.06$. Repeating the process a realization with all 30 sites occupied would typically occur in $\sim10$ s.

The main graph in Fig. \ref{fig3}(a) shows the single atom loading efficiency as a function of $\Delta_c$, as well as the probability of loading zero or two atoms.
The MOT parameters and the duration of the collision pulse were fixed at the values used in the inset.
We have adjusted the collision power at each frequency in order to maximize the probability of loading one atom at the end of the
collision process. 
As noted above, a critical feature for efficient loading is a long lifetime of the single atom once it has been prepared. In order to monitor this, we prepared single atoms, exposed them to a collision pulse of duration $3.5$ s, and measured the probability that the atom remained trapped. Fig. \ref{fig3}(b) shows this survival probability for each set of collision pulse parameters. We see that the lifetime of a single atom decreases when the collision beam frequency is
close to resonance (small $\Delta_c$). We ascribe this to the
increased heating from radiation pressure by the collision beam at small detunings, leading to a higher equilibrium temperature.
Although the equilibrium temperature remains below the trap depth there is a finite probability that the atom will occupy the high energy tail
of the distribution and thus be lost. The reduction of the lifetime  at small detunings leads to a drop in the single atom loading efficiency.

To understand the reduction in single atom loading probability observed for large detunings in Fig. \ref{fig3}(a) we measured the time
evolution of pairs in a similar way to Fig. \ref{fig2} but with collision beam parameters identical to those of the $\Delta_c=185$ MHz point in Fig. \ref{fig3}(a).
In most of the cases, both collision partners were lost together
as the energy released in each collision event is enough for both atoms to escape.

%
%%%%%%%%%%%%% figure 4 %%%%%%%%%%%%%%
\begin{figure}
 \epsfig{figure=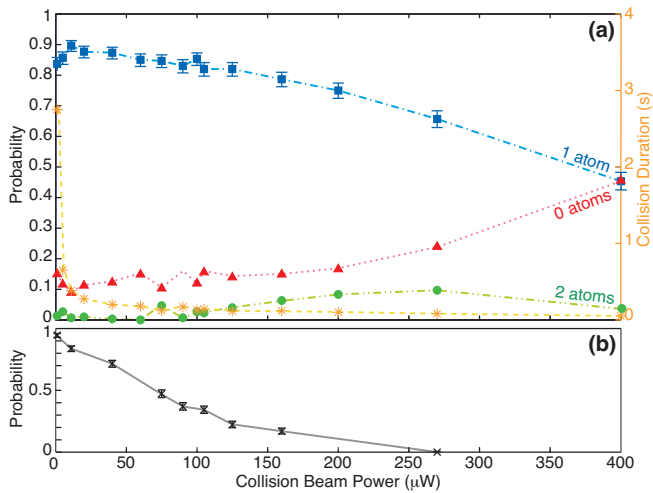, width=1\columnwidth}
\caption{ (Color online) (a): Loading probabilities as a function of the collision beam power.
The meaning of the symbols are as in previous figures. 
The duration of the collision pulse is adjusted for each power and
is plotted as orange stars, using the time axis on the right of the graph. (b): Single atom survival probability after a collision pulse of $3.5$ s.
 }
\label{fig4}
\end{figure}
%%%%%%%%%%%%%%%%%%%%%%%%%%%%%%%
%
%
We have also studied how the single atom loading efficiency evolves as a function of the collision beam power.
The result is plotted in Fig. \ref{fig4} along with the probability of loading zero or two atoms. For each value of the power, we have adjusted the duration of the pulse
in order to maximize the probability of loading one atom in the microtrap. The orange stars and orange right axis in Fig. \ref{fig4}(a) show the duration of the applied pulse.
The other parameters are the same as in Fig. \ref{fig2}.
The peak probability is reached for a collision power of $11$ $\mu$W and a collision pulse duration of $385$ ms, consistent with the previous data.
At low powers, the single atom lifetime increases (being $\sim63$ s for $1$ $\mu$W of collision beam power)
but the optimized  duration of the collision pulse increases to $2.75$ s. During this time, some pairs are
lost due to processes not caused by the collision beam; in particular light assisted collision induced by the MOT cooling beams.
We measured the evolution of pairs in the presence of only the MOT cooling beams and found a decay time of $\sim4$ s
(comparable to the $2.75$ s). In each event both atoms are lost. Pair loss is indeed expected from the fact that the
MOT beams are $\sim3$ GHz red detuned from the $D2$ line for atoms colliding in the $F=2$ ground state \cite{DePueMT,Schlosser}.
We attribute the reduction in the loading efficiency at high collision beam powers to the reduced single atom lifetime (see Fig. \ref{fig4}(b)) from increased heating from radiation pressure.

We estimate the contributions of a series of processes to the unsuccessful $9\%$ of the realizations in the following way:
our single atom detection efficiency of $99.5\%$ gives a contribution of $\sim 0.5\%$ of the realizations where a single atom
is loaded but not detected.
By running a Monte-Carlo simulation for the loading process as described in \cite{andersen}, using the measured single atom lifetime of
$\sim22$ s, we estimate the single atom loss from imperfect vacuum and finite temperature to contribute $\sim1.5\%$. We attribute the remaining $\sim7\%$ to pair losses.
Based on the pair decay time caused by the MOT beams only, and that of Fig. \ref{fig2},
we estimate that $\sim1.7\%$ are due to the inelastic collisions induced by the MOT beams during the collision process. Of the
remaining $\sim5.3\%$, the simulation in Fig. \ref{fig2} predicts that $3.6\%$ comes from
collision events that occur at sufficiently high $E_p$ leading to the escape of both partners.
The remaining contribution could come from inelastic collisions that cause the atom to change hyperfine state.

In future work it could be interesting to explore the possibilities of using other atomic species such as the $^{87}$Rb isotope or Cs. Their larger hyperfine splittings may help to suppress the unwanted effects from the MOT cooling light and hyperfine changing collisions. Eliminating these effects may lead to loading efficiencies in excess of $95\%$. 

In conclusion, we have demonstrated a single atom loading efficiency of $91\%$ in a red detuned optical microtrap.
We have used blue detuned light assisted collisions in combination with laser cooling in order to eject atoms from the trap one by one.
A simple model of the process quantitatively agrees with the experimental measurements.
The loading can be completed in $542$ ms potentially allowing a fully-loaded array of 30 sites in $\sim10$ s \cite{quantumRegister}.

This work is supported by NZ-FRST Contract No. NERF-UOOX0703 and UORG. T. W.'s work is funded by the National Science Foundation and the AFOSR Light-Matter Interfaces MURI.

\end{document}